\def\del{\partial}

\def\abs#1{{\left|{#1}\right|}}
\def\vev#1{\langle #1 \rangle}

\def\del{\partial}
\def\dslash{\del\kern-0.55em\raise 0.14ex\hbox{/}}

\def\Fig{Fig.~\the\figno\nfig}
\newcommand{\PRD}[3]{Phys. Rev. {\bf D{#1}} (19{#2}) {#3}}
\newcommand{\PRDM}[3]{Phys. Rev. {\bf D{#1}} (20{#2}) {#3}}
\newcommand{\PRL}[3]{Phys. Rev. Lett. {\bf {#1}} (19{#2}) {#3}}
\newcommand{\NPB}[3]{Nucl. Phys. {\bf B{#1}} (19{#2}) {#3}}
\newcommand{\PLB}[3]{Phys. Lett. {\bf B{#1}} (19{#2}) {#3}}
\newcommand{\PLBM}[3]{Phys. Lett. {\bf B{#1}} (20{#2}) {#3}}

\newcommand{\ANN}[3]{Ann. Phys. {\bf {#1}} (19{#2}) {#3}}

\newcommand{\RPP}[3]{Rep. Prog. Phys. {\bf {#1}} (19{#2}) {#3}}
\newcommand{\PRS}[3]{Proc. R. Soc. London. {\bf A{#1}} (19{#2}) {#3}}
\newcommand{\SJN}[3]{Sov. J. Nucl. Phys. {\bf {#1}} (19{#2}) {#3}}

\documentstyle[11pt,epsf]{article}
\begin{document}
\date{} 
\begin{flushright}
{KOBE-TH-00-07}\\
{MIT-CTP-3038}\\
\end{flushright}
\vspace{1cm}
\begin{center}
{\Large {\bf SUSY BREAKING THROUGH COMPACTIFICATION}
\footnote{Talk given at XXXth International Conference on High Energy
Physics, July 27-August 2, 2000, Osaka, Japan}}
\vskip0.5truein
{\large Makoto Sakamoto}$^{(a)}$
\footnote{E-mail:{\tt sakamoto@phys.sci.kobe-u.ac.jp}},
{\large Motoi Tachibana}$^{(b)}$ 
\footnote{E-mail:{\tt motoi@@mitlns.mit.edu}~~~on leave from
Yukawa Institute for Theoretical Physics, 
Kyoto University, Kyoto 606-8502, Japan}\\
\vspace*{2mm}and \vspace*{2mm}\\
{\large Kazunori Takenaga}$^{(c)}$
\footnote{E-mail:{\tt takenaga@synge.stp.dias.ie}}
\vskip0.2truein
\centerline{$^{(a)}$ {\it Department of Physics,
Kobe University, Rokkodai, Nada, Kobe 657-8501, Japan }}
\vspace*{2mm}
\centerline{$^{(b)}$ {\it Center for Theoretical Physics, 
Massachusetts Institute of Technology,
Cambridge, MA 02139, USA}}
\vspace*{2mm}
\centerline{$^{(c)}$ {\it Dublin Institute for Advanced Studies, 
10 Burlington Road, Dublin 4, Ireland}}
\end{center}
\vskip0.5truein
\centerline{\bf Abstract}
\vskip0.3truein
We propose a new mechanism of spontaneous supersymmetry breaking. 
The existence of extra dimensions with nontrivial topology plays an
important role. We investigate new features resulting from this mechanism.
One noteworthy feature is that there exists a phase
in which the translational invariance for the compactified directions is
broken spontaneously.
The mechanism we propose also yields quite different vacuum structures
for models with different global symmetries.
\vskip0.13truein
\baselineskip=0.5 truein plus 2pt minus 1pt
\baselineskip=18pt
\addtolength{\parindent}{2pt}
It is thought that (super)string
theories \cite{gsw} (and/or more fundamental theories, such as M-theory)
are plausible candidates
to describe physics at the Planck scale. In general these theories are
defined in more than $4$ space-time dimensions because of the
consistency of the theories. \par
In the region of sufficiently low energy, however, it is known
that our space-time is $4$ dimensions, so that extra dimensions must be
compactified by some mechanism \cite{kaluza}, and
supersymmetry (SUSY) that
is usually possessed by these theories has to be 
broken \cite{antoniadis} because it is not
observed in the low energy region.
At this stage, the mechanism of compactification 
and SUSY breaking are not fully understood.
\par
It may be interesting to consider the quantum field theory in space-time
with some of the space coordinates being 
multiply-connected. This is because
such a study may shed new light on unanswered questions
concerning the (supersymmetric) standard model and/or elucidate
some new dynamics which are worth seeking and understanding
new physics beyond it.
Actually, it has been reported that flavour-blind SUSY breaking
terms can be induced through compactification by taking account 
for possible topological effects of
the multiply-connected space \cite{ss,fi2,takenaga}.
Therefore, it is important to investigate the physics
possessed by the quantum field theory considered in such a space-time.
\par
We consider supersymmetric field theories in
space-time with one of the space coordinates being
compactified. One has to specify the
boundary conditions of fields for the compactified direction
when the compactified space is
multiply-connected. In contrast to finite
temperature field theory, we
do not know {\it a priori} what they should be.
We shall relax the conventional periodic boundary condition
to allow nontrivial boundary
conditions for the fields \cite{hosotani,isham}.
%
%
One of the possible physical origins of the nontrivial boundary condition
comes from the quantization ambiguity of the theory in multiply-connected 
space. An undetermined parameter, which appears as the consequence of the
ambiguity, twists the boundary condition of the 
field\footnote{In a SUSY quantum mechanical system on a 
multiply-connected space, it has been shown that the 
ground state wave function is actually twisted by the parameter\cite{take2}}. 
\par
As a noteworthy consequence of our mechanism,
there appears a nontrivial phase structure with respect to
the size of the compactified space. Namely, 
the translational invariance for the compactified direction
is broken spontaneously \footnote{Models which cause the spontaneous
breakdown of the translational invariance 
are discussed \cite{stt0}.} when the size of the compactified
space exceeds a certain critical value. The curious
vacuum structures resulting from the
mechanism also have an influence on the mass spectrum of the theory.
Depending on the size of the compactified space,
we have a mass spectrum full of variety.
The mass spectrum includes Nambu-Goldstone bosons (fermions)
corresponding to the breakdown of the global (super) symmetries of the theory.
\par
Let us discuss a basic idea of our SUSY-breaking mechanism briefly.
Let $W(\Phi)$ be a superpotential consisting of the
chiral superfields $\Phi_j$. The scalar potential is then given by
\begin{equation}
V(A)=\sum_j\abs{F_j}^2=\sum_j\abs{\frac{\del W(A)}{\del A_j}}^2,
\label{supot}
\end{equation}
where $A_j(F_j)$ denotes the lowest (highest) component of $\Phi_j$.
Supersymmetry would be unbroken if there exist
solutions to the ${\it F}$-term conditions
\begin{equation}
-F_j^*=\frac{\del W(A)}{\del A_j}\bigg|_{A_k = \bar{A}_k} = 0 \qquad
{\rm for~~all} \quad j,
\label{fterm1}
\end{equation}
since such solutions would lead to $V(\bar{A})=0$. Our idea for
supersymmetry breaking is simple: We impose nontrivial boundary
conditions on the superfields for the compactified direction.
They must be consistent with the single valueness of the Lagrangian but 
inconsistent with the ${\it F}$-term conditions
(\ref{fterm1}). Then, no solution to the ${\it F}$-term conditions
will be realized as a vacuum configuration of the model.
Thus, we expect that $V(\langle A \rangle) > 0$ (because
$\langle A_j \rangle \neq \bar{A_j}$) and that supersymmetry
is broken spontaneously. 
\par
In order to realize such a mechanism, let us
consider a theory in 
one of the space coordinates, say $y$, being compactified on a circle
$S^1$ whose radius is $R$. Since $S^1$ is multiply-connected, we must
specify boundary conditions for the $S^1$ direction. Let us
impose nontrivial boundary conditions on superfields defined by
\begin{equation}
\Phi_j(x^{\mu}, y+2\pi R) = e^{2\pi i \alpha_j}\Phi_j(x^{\mu}, y).
\label{bccond}
\end{equation}
The phase $\alpha_j$ should be chosen such that the Lagrangian density
is
single-valued, {\it i.e.}
\begin{equation}
{\cal L}(x^{\mu}, y+2\pi R) = {\cal L}(x^{\mu}, y).
\label{lagra}
\end{equation}
In other words, the phase has to be the symmetry degrees of 
freedom possessed by the theory. 
Suppose that $\bar{A_j}$, which is a
solution to the ${\it F}$-term conditions, is a nonzero
constant for some $j$. It is easy to see that if
\begin{equation}
e^{2\pi i \alpha_j} \neq 1,
\label{nontbc}
\end{equation}
then the vacuum expectation value $\vev{A_j}$ is strictly forbidden
to take the nonzero constant value $\bar{A_j}$, because this is
inconsistent with the boundary condition (\ref{bccond}).
In this way, our idea is realized by the mechanism that
the nontrivial boundary conditions imposed
on the fields play the role of preventing
the vacuum expectation values of the fields 
from being solutions to the ${\it F}$-term conditions.
\par
Note that the above result does not always lead us to the
conclusion that $\vev{A_j}= 0$, though it is always consistent
with (\ref{bccond}).
Certainly, if the translational invariance for the $S^1$ direction is
not broken, $\vev{A_j}$ has to vanish because
of (\ref{bccond}) with (\ref{nontbc}).
If the translational invariance for the $S^1$ direction
is broken, however, vacuum expectation values will
no longer be constants, and some of the $\vev{A_j}$ can depend on
the coordinate of the compactified space as an energetically
favourable configuration. One should include
the contributions from the kinetic terms of the scalar fields in addition
to the scalar potential in order to find the true vacuum configuration.
\par
Let us illustrate how to 
realize the above idea. For simplicity, here
we consider $j=1$ case. First we expand the complex scalar field 
$A(x^{\mu}, y)$ in Fourier modes
\begin{equation}
A(x^{\mu}, y) = \frac{1}{\sqrt{2\pi R}}\sum^{\infty}_{n=-\infty}
A^{(n+\alpha)}(x^{\mu})e^{i\frac{n+\alpha}{R}y}.
\label{fourier}
\end{equation}
Note here that due to the twisted boundary condition (\ref{bccond})
there are no zero modes. Moreover, we introduce  
dimensionally reduced \lq \lq effective"  potential defined by
\begin{eqnarray}
{\cal V}_{eff}[A] &\equiv& \int^{2\pi R}_0 dy \Bigl[ |\partial_y A|^2 
+V(A)\Bigr] \nonumber \\
&=& \sum^{\infty}_{n=-\infty}\Bigl[\Bigl(\frac{n+\alpha}{R}\Bigr)^2
-\mu^2\Bigr]|A^{(n+\alpha)}|^2+\cdots \nonumber \\
&=& \Bigl[\Bigl(\frac{\alpha}{R}\Bigr)^2-\mu^2\Bigr]|A^{(\alpha)}|^2+\cdots,
\label{effective}
\end{eqnarray}
where $A^{(\alpha)}$ is the lowest mode and the dots 
denote higher order modes and terms including 
the quartic term of $A^{(\alpha)}$.
The $\mu^2$ term comes from the potential $V(A)$ which is assumed
to include a negative mass squared term.
From the potential it is easy to find that
if $R < R^* \equiv \frac{\alpha}{\mu}$, the curvature at the origin of the
potential is positive so that $\langle A(x^{\mu},y) \rangle$
vanishes and $\langle A(x^{\mu},y) \rangle =0$ becomes a stable vacuum. 
Thus, in this case, the translational invariance for the $S^1$ 
direction is not broken. On the other hand,
if $R > R^*$, the curvature becomes negative and the potential
becomes the double-well type. Then the vacuum expectation value of
$A(x^{\mu},y)$ does not vanish but rather has 
$y$-dependence. In this case, the
translational invariance for the $S^1$ direction
is spontaneously broken. See Fig.1 below.
The vacuum configuration for $R > R^*$ is not
a solution of the ${\it F}$-term condition, so the SUSY is 
spontaneously broken.
\begin{center}
\leavevmode{\epsfxsize=10cm\epsffile{eff.eps}}\\
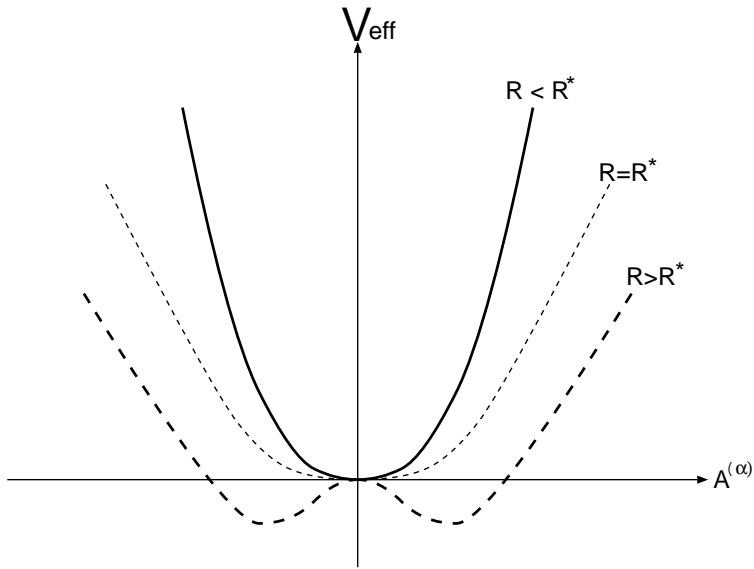
Fig.1 The \lq \lq effective" potential ${\cal V}_{eff}(A^{(\alpha)})$.
\end{center}
\smallskip
\par
Finally we should comment on the Scherk-Schwarz (S-S) mechanism
\cite{ss,fi2}. One might impose nontrivial boundary conditions
associated with a $U(1)_R$ symmetry. In that case, bosonic components
of superfields may satisfy boundary conditions different from the
fermionic ones. A crucial difference between the Scherk-Schwarz
mechanism and ours is that the breaking \`a la Scherk-Schwarz
is {\it explicit}, rather than spontaneous, at the level of global
supersymmetry. Another difference is that the S-S mechanism will work
for any choice of superpotentials, while ours will not.
The boundary conditions (\ref{bccond})
we consider in this paper are consistent with supersymmetry.
When a space is multiply-connected, and hence has nontrivial
topology, the configuration space can also have nontrivial
topology. Then, it turns out that undetermined parameters
like the $\alpha_j$ in (\ref{bccond}) inevitably appear as the
ambiguity in quantizing the theory on such a space, and the
boundary conditions are twisted in accordance with the parameters
\cite{rs}. Therefore, even though we
impose nontrivial boundary conditions on the fields here, these
boundary conditions have a firm foundation
that is clarified when one studies the
theory on multiply-connected space.
\par
It may be interesting to ask how our mechanism works on more complex
manifolds, such as a torus. We expect that there will appear rich
phase structures, depending on the size of their compactified spaces
and on how we impose nontrivial boundary conditions on superfields.
In the previous papers \cite{stt1,stt2}, we studied the $Z_2$ and $U(1)$
models and found that the vacuum structures of the models are
quite different from each other.
We think the rich vacuum structure is a general 
feature of our mechanism. We can also
study models with non-abelian global symmetries, such as $SU(N)$ 
and $O(N)$\cite{taishou}.
In addition, it is important to study gauge theories
and to see how our mechanism works and what new dynamics are hidden in them.
It may also be interesting to investigate how partial SUSY
breaking occurs in  gauge theories in connection with the
well-known BPS objects.\\ 
\\
{\large{\bf Acknowledgments}}\\
This work was supported in part by Grant-In-Aid for 
Scientific Research (No.12640275) from the Ministry of 
Education, Science, and Culture, Japan (M.S)
and by a Grant-in-Aid for Scientific Research, Grant No.~3666 (M.T).
K. T. would like to thank DIAS, Niels Bohr Institute and
INFN, Sezione di Pisa for warm hospitality. 

\end{document}